\documentclass[twocolumn]{aastex63}  

\usepackage{verbatim}
\usepackage{float}
\usepackage{graphicx}
\usepackage{epstopdf}
\usepackage{subfigure}
\usepackage{CJK}
\usepackage{amsmath}
\usepackage{booktabs}
\usepackage{url}
\usepackage{amssymb}
\usepackage{multirow}
\usepackage{bm}

\makeatletter

\newcommand{\Rmnum}[1]{\expandafter\@slowromancap\romannumeral #1@}
\makeatother

\shorttitle{}
\shortauthors{}

\begin{document}

\begin{CJK*}{UTF8}{gbsn}

\title{Patterns of X-ray and $\gamma$-ray Flares: from Blazar to Maganetar and Sun}

\author{Haiyun Zhang (张海云)}
\affiliation{Department of Mathematics and Statistics, Yunnan University, Kunming 650091, China;}

\author{Dahai Yan (闫大海)}
\affiliation{Department of Astronomy, Key Laboratory of Astroparticle Physics of Yunnan Province, Yunnan University, Kunming 650091, China;}
\email{yandahai@ynu.edu.cn}

\author{Lijuan Dong (董丽娟)}
\affiliation{Department of Astronomy, Key Laboratory of Astroparticle Physics of Yunnan Province, Yunnan University, Kunming 650091, China;}

\author{Ping Zhang (张平)}
\affiliation{School of Sciences，Wuhan University of Science and Technology
WuHan, HuBei, China;}

\author{Ruijing Tang (唐睿婧)}
\affiliation{Department of Physical Science and Engineering, Beijing Jiaotong University, Beijing 100044, Peopleʼs Republic of China;}

\author{Jianeng,Zhou (周佳能)}
\affiliation{Shanghai Astronomical Observatory, Chinese Academy of Sciences, 80 Nandan Road, Shanghai 200030, People's Republic of China}

\author{Lian Tao (陶炼)}
\affiliation{Key Laboratory of Particle Astrophysics, Institute of High Energy Physics, Chinese Academy of Sciences, Beijing 100049, Peopleʼs Republic of China;}

\author{Li Zhang (张力)}
\affiliation{Department of Astronomy, Key Laboratory of Astroparticle Physics of Yunnan Province, Yunnan University, Kunming 650091, China;}
\email{lizhang@ynu.edu.cn}

\author{Niansheng Tang (唐年胜)}
\affiliation{Department of Mathematics and Statistics, Yunnan University, Kunming 650091, China;}
\email{nstang@ynu.edu.cn}

\begin{abstract}
Using Gaussian process methods, we analyzed the light curves of three extreme solar X-ray flares observed by the RHESSI satellite. Their variability characteristics were then compared with those of HXMT-HE X-ray burst (XRB; in SGR 1935+2154) associated with fast radio burst (FRB) 200428 and blazar $\gamma$-ray giant flares, to investigate the origins of these extreme flaring events.
The variability patterns of the solar X-ray flares follow the stochastically driven damped simple harmonic oscillator (SHO) model.
The derived timescales $t_{\rm B\underline{~} steep}$ and $t_{\rm B\underline{~} flat}$ (corresponding to PSD breaks) are in the range of 4-7 s and 16-53 s, respectively.
The FRB-associated HXMT-HE burst has a $Q$ value near 0.3, matching those of the solar flares occurred on 23 July 2002 (flare 1) and 3 November 2003 (flare 2).
By contrast, blazar $\gamma$-ray giant flares show Q $>$ 0.3, similar to the solar flare occurred on 25 February 2014 (flare 3).
We proposed that the critically damped state of the system may be the condition triggering the association between the XRB in SGR 1935+2154 and the FRB.
In this scenario, the critical damping $Q$ value of the system is around 0.3, not the theoretical 0.5.
The similarity in $Q$ values might imply that the FRB-associated HXMT-HE XRB and solar X-ray flares 1 $\&$ 2 share comparable dynamic behavior, while blazar $\gamma$-ray flares and solar X-ray flare 3 exhibit another distinct but similar dynamic behavior.
Like solar X-ray flares, these extreme flares may all be related to the magnetic reconnection process. 
\end{abstract} 

\keywords{Burst astrophysics (187), Solar X-ray flares (1816), Time series analysis (1916), Gaussian Processes regression (1930)}

\section{Introduction} \label{sec:intro}

Flaring phenomena in different astronomical objects, such as solar flares, magnetar X-ray bursts (XRBs), $\gamma$-ray bursts and large flares from active galactic nuclei (AGNs), typically involve the processes of energy release, particle acceleration, and magnetic activity, with their physical environments, underlying mechanisms and radiation signatures potentially differing or exhibiting similarities in certain aspects.
Studying the time-domain characteristics of different flare events allows us to gain deeper insights into involved physical processes in extreme astrophysical conditions.

Blazars are a distinct subclass of AGNs, characterized by their relativistic jets pointing directly toward the Earth \citep[e.g.,][]{2015ApJ...810...14A}. 
They emit radiation across entire electromagnetic spectrum. Blazars typically exhibit strong variability, which results in prominent time-domain characteristics in multi-wavelength observations.
To study such variability, the Gaussian Process \citep[GP;][]{2021ApJ...919...58Z,2022ApJ...930..157Z,2023ApJ...944..103Z} method has been increasingly applied in recent years.
GP is a powerful data analysis tool based on the theory of probability statistics and Bayesian analysis techniques, providing a flexible, non-parametric framework for modeling time series data. In GP, the kernel function encodes assumptions about the correlation structure, allowing us to capture both smooth trends and stochastic fluctuations.
Results from the analyzed studies, obtained through GP modeling, indicate that the long-term variability of Blazars follows the damped random walk (DRW) model, with the characteristic timescales associated with the thermal instability timescale of the accretion disk \citep{2022ApJ...930..157Z,2023ApJ...944..103Z}.
As prominent blazars, both 3C 454.3 and 3C 279 exhibit $\gamma$-ray flaring events whose light curves were suggested to follow the stochastically driven damped simple harmonic oscillator (SHO) model in the overdamped regime \citep{2025MNRAS.540.3790Z}.
This variability behavior is typically attributed to internal processes within the innermost regions of their relativistic jets \citep[e.g.,][]{2016ApJ...824L..20A,2018ApJ...854L..26S}.

Magnetars are a type of neutron stars with extremely strong magnetic fields ($10^{14}-10^{16}$ G; see \cite{2017ARA&A..55..261K} for a review),
whose intense magnetic activity can trigger multi-band radiation bursts, e.g., XRBs, radio bursts (may include fast radio bursts, FRBs), and $\gamma$-ray bursts \citep{2015RPPh...78k6901T}.
Studies suggest that the energy release from magnetar bursts could be linked to direct magnetic reconnection in the magnetosphere, driven by the intense magnetic field \citep[e.g.,][]{2013ApJ...762...13B,2015SSRv..191..315M}. 
Alternatively, these flares might be triggered by shear waves released during crustal rupture, which transmit energy to the magnetosphere through Alfv$\acute{\rm e}$n waves, initiating secondary magnetic reconnection \citep[e.g.,][]{2015ApJ...815...25L,2020ApJ...900L..21Y}.
SGR J1935+2154 is a Galactic magnetar, of which hundreds of XRBs 
were recorded by different detectors, e.g., Insight-HXMT \citep{2021NatAs...5..378L}, Neutron Star Interior Composition Explorer \citep[NICER;][]{2020ApJ...900L..21Y}.
A non-thermal XRB of SGR J1935+2154 on 2020 April 28th has been discovered to be accompanied by emission of the FRB 200428 \citep{2020Natur.587...54C,2021NatAs...5..378L}, suggesting the non-thermal XRB and FRB 200428 share the same physical origin in an explosive event from SGR J1935+2154.
For XRBs in SGR J1935+2154, \cite{2024ApJ...971...26T} used GP method to characterize their variability.
They found that both normal XRBs and FRB-associated HXMT XRB generally show variability consistent with the SHO patern, the derived timescales can put constraints on the production mechanisms of XRBs.

Solar flares are explosive phenomena in the solar atmosphere, which can emit electromagnetic emission extending to $\gamma$-rays within a short time.
Extensive research has been conducted on particle acceleration, radiation mechanisms, and emission regions in solar flares \citep[e.g.,][]{2011SSRv..159..107H,2020NatAs...4.1140C,2024arXiv241219586S}.
It is generally agreed that energy release is usually attributed to magnetic reconnection occurring on the current sheet \citep[e.g.,][]{2011LRSP....8....1C,2017LRSP...14....2B,2018ApJ...869..182S,2020NatAs...4.1140C}.
Research indicates that the light curves of hard X-rays ($>20$ keV) display fast pulsating structures with periods ranging from milliseconds to seconds, closely related to the acceleration of high-energy electrons induced by magnetic reconnection, with non-thermal bremsstrahlung as the primary radiation mechanism \citep[e.g.,][]{2011SSRv..159..107H,2019ApJ...871..225K,2020ApJ...894..158K}.

Building on the analysis of the variability patterns of blazar $\gamma$-ray flares and magnetar XRBs, we would like to explore the variability of the solar flares by GP method in this work.
Furthermore, we will compare the burst properties of solar X-ray flares, magnetar XRBs, and $\gamma$-ray blazar outbursts, focusing on investigating the specific mechanisms driving high-energy flare phenomena across these different types of sources.
The structure of this paper is as follows: Section \ref{sec:method} describes the construction of the hard X-ray light curves of solar flares, as well as the models and methods used to simulate the light curves. 
Section \ref{sec:results} displays the fitting results for solar X-ray flares by three different kernels, and compares the variability characteristics of solar X-ray flares, magnetar XRBs, and $\gamma$-ray outbursts of blazars.
In Section \ref{sec:discussion}, we discuss the possible origins of the three types of bursts and the potential issues associated with them, based on the results.

\section{Method and data processing} \label{sec:method}
The Reuven Ramaty High Energy Solar Spectroscopic Imager (RHESSI) is a NASA Small Explorer Mission \cite{2002SoPh..210....3L}. It is capable of performing both imaging and spectroscopic analyses across a broad energy spectrum, ranging from soft X-rays to $\gamma$-rays (3 keV to 17 MeV). The highest energy resolution is as fine as 1 keV. 
In this study, we extract hard X-ray lightcurves using the SSWIDL software, based on the RHESSI Observing Summary Data. The summary lightcurves provide count rates in nine pre-selected energy bands with a temporal resolution of 4 s. Heuristic corrections are applied to remove the effects of attenuator and decimation state changes; these corrections are sufficient for timing and variability studies, although they do not provide absolute photon fluxes which are generally required only for detailed spectral analysis.
For our analysis, we focus on high-energy bands where the effect of pile-up is less pronounced, and therefore do not explicitly consider attenuator status, which mainly affects low-energy ($<$40 keV) observations. All light curves were generated using detectors 1, 3, 4, 5, 6, and 9, which are known for their stable performance and low noise levels. 
We conduct a detailed analysis of the flare light curves within the GP framework \citep{2023ARA&A..61..329A}, focusing on the near-peak phase of the solar flares, which approximately corresponds to the impulsive phase (highlighted by the gold-shaded region in Figure~\ref{fig:LC}). 

GP can be regarded as a distribution over functions, fully specified by a mean function and a covariance (kernel) function. 
Instead of fitting a fixed functional form, GP places a prior directly on the space of possible functions, and the kernel determines how data points are correlated in time. 
This property makes GP particularly suitable for irregularly sampled and noisy astrophysical light curves. 
In addition, GP is capable of characterizing fast variabilities, such as those frequently associated with quasi-periodic pulsations (QPPs) which are commonly observed during the impulsive phase of solar flares.
In our modeling, the mean function of the GP is chosen to be a constant equal to the average count rate of the light curve for each flare.
The kernels act as key components of the GP framework. They are mathematical models whose primary function is to quantify the similarity between data points in time, thereby governing the structure and smoothness of the resulting light-curve models.
Following empirical practice, we opt for three kernel functions to fit the light curves, i.e., DRW, SHO, and Mat$\acute{\rm e}$rn$-3/2$ which are used in our previous works \citep{2021ApJ...919...58Z,2022ApJ...930..157Z,2023ApJ...944..103Z}.

The DRW process can be described by the following first-order stochastic differential equation \citep{2009ApJ...698..895K}:
\begin{equation}
  \left[\frac{d}{dt}+\frac{1}{\tau_{\rm DRW}}\right]y(t)=\sigma_{\rm DRW}\epsilon(t),
    \label{DRW}
\end{equation}
where $\sigma_{\rm DRW}$ represents the amplitude term, $\tau_{\rm DRW}$ is correlation decay timescale or damping timescale, and $\epsilon(t)$ represents a white noise process.
The kernel function of the DRW is written as:
\begin{equation}
    k_{\text{DRW}}(t_{nm})=2\sigma_{\rm DRW}^2 e^{-\frac{t_{nm}}{\tau_{\rm DRW}}}. \label{DRW2}
\end{equation}
The power spectral density (PSD) of this process is performing as a broken power-law form \citep[e.g.,][]{2010ApJ...721.1014M,2021Sci...373..789B}:
\begin{equation}
    S(\omega)=\sqrt{\frac{8}{\pi}}\frac{\sigma_{\rm DRW}^{2}\tau_{\rm DRW}}{1+(\omega\tau_{\rm DRW})^2}\ ,\label{DRW_PSD}
\end{equation}
with the spectral index transiting from -2 to 0\footnote{The change of spectrum index mentioned in this paper refers to the change from high frequencies to low frequencies.}. 
The DRW kernel corresponds to the Ornstein-Uhlenbeck (OU) process \citep{PhysRev.36.823} in physics. It is typically used to capture random fluctuations that gradually decay.

The form for a SHO process driven by a random (white noise) forcing $\epsilon(t)$ 
is:
\begin{equation}
    \left[ \frac{d^2}{dt^2} + \frac{\omega_0}{Q} \frac{d}{dt} + \omega_0^2 \right] y(t) = \epsilon(t),\label{SHO_t}
\end{equation}
where $\omega_0$ is the natural frequency of the undamped oscillator, and $Q$ is its quality factor of the oscillator.
The SHO kernel function is relatively sophisticated, and we refer the reader to \cite{2017AJ....154..220F} for a more detailed description.
The PSD of this process is
\begin{equation}
    S(\omega) = \sqrt{\frac{2}{\pi}} \frac{S_0 \omega_0^4}{(\omega^2 - \omega_0^2)^2 + \frac{\omega_0^2 \omega^2}{Q^2}}, \label{SHO}
\end{equation}
where $S(\omega_0)=\sqrt{2/\pi} S_0 Q^2$.
The SHO represents a more complex model that can be categorized into three regimes according to the value of the oscillation quality factor $Q$.
When $Q<0.5$, it falls in the overdamped regime. In this regime, the PSD matches the DRW PSD at low frequencies, i.e., the spectral index transitions from -2 to 0 (characteristic timescale denoted as $t_{\rm B\underline{~} flat}$), and the spectral index drops to -4 at high frequencies (where the timescale $t_{\rm B\underline{~} steep}$ corresponds to the transition of the spectral index from -4 to -2).
The two characteristic timescales can be calculated as:
\begin{equation}\label{eq:t_flat}
t_{\rm B\underline{~} flat}={\rm max}\ (\frac{1}{\omega_{0}},\frac{1}{\omega_{0}Q}\sqrt{1-2Q^{2}})\;,
\end{equation}
\begin{equation}\label{eq:t_steep}
t_{\rm B\underline{~} steep}={\rm min}\ (\frac{1}{\omega_{0}},\frac{Q}{\omega_{0}\sqrt{1-2Q^{2}}})\;.
\end{equation}
When $Q>0.5$, SHO falls in the underdamped regime, where its PSD shows a peak \citep[e.g.,][]{2021ApJ...907..105Y,2021ApJ...919...58Z,2025MNRAS.537.2380Z}, representing a quasi-periodic oscillation (QPO) signal.
When $Q=0.5$, it falls in the critically damped regime, with the spectral index of PSD tending to transform from -4 directly to 0 \citep[refer to Figure~\ref{fig:sim_PSD} for a more intuitive representation;][]{2017MNRAS.470.3027K,2019PASP..131f3001M}.
The SHO model is most commonly used to describe periodic behaviors in time series.
Actually, depending on the regime, it can capture distinct characteristics of the variability.

When setting the smoothness parameter $l$ (half-integer) of the Mat$\acute{\rm e}$rn class kernels to 3$/$2, we get Mat$\acute{\rm e}$rn$-3/2$. 
This process can be described by a second-order stochastic differential equation \citep{2012Kalman}:
\begin{equation}
    \left[ \frac{d^2}{dt^2} + \frac{2\sqrt{3}}{\rho}\frac{d}{dt} + \frac{3}{\rho^2} \right] y(t) = \epsilon(t).\label{matern_t}
\end{equation}
$\epsilon(t)$ represents white noise process, $\rho$ is the same as $\tau_{\rm DRW}$ in DRW model, representing correlation decay timescale.
The  kernel function of Mat$\acute{\rm e}$rn$-3/2$ is:
\begin{equation}
  k (t_{nm})=\sigma^{2}(1+\frac{\sqrt{3}t_{nm}}{\rho}){\rm exp}(-\frac{\sqrt{3}t_{nm}}{\rho})\ ,
    \label{maternkernel}
\end{equation}
where $\sigma$ represents the amplitude terms.
The corresponding PSD can be written as \citep{2006gpml.book.....R,2012Kalman}:
\begin{equation}
  S (\omega)\propto (\frac{3}{\rho^2}+\omega^2)^{-2},
    \label{maternpsd}
\end{equation}
with its spectral index transforming from -4 to 0.
Compared to the DRW (equal to the  Mat$\acute{\rm e}$rn class kernel with $l=1/2$) model, the Mat$\acute{\rm e}$rn$-3/2$ model produces curves that are smoother at high frequencies, and its smoothness and correlation can be adjusted by the parameter $\rho$.

we use a tool of {\it Celerite} \citep{2017AJ....154..220F} to perform GP fitting time series.
Markov chain Monte Carlo (MCMC) technology is used when implementing the light curve fitting.
We implement it through the package emcee \footnote{\url{https://github.com/dfm/emcee}} \citep{2013PASP..125..306F}, taking the last 30,000 steps out of 50,000 total MCMC steps as effective samples.
Actually, before running MCMC, we choose the initial parameter values through calculating the maximum likelihood.
Then, we can construct the final PSD by taking the median of the 30,000 PSD values corresponding to each frequency, and plot 30,000 groups of parameters as a distribution, i.e., a posterior distribution of the model parameters.

\section{results} \label{sec:results}
\subsection{results of the solar X-ray flares}

\begin{figure}
    \centering
    {\includegraphics[width=1\linewidth]{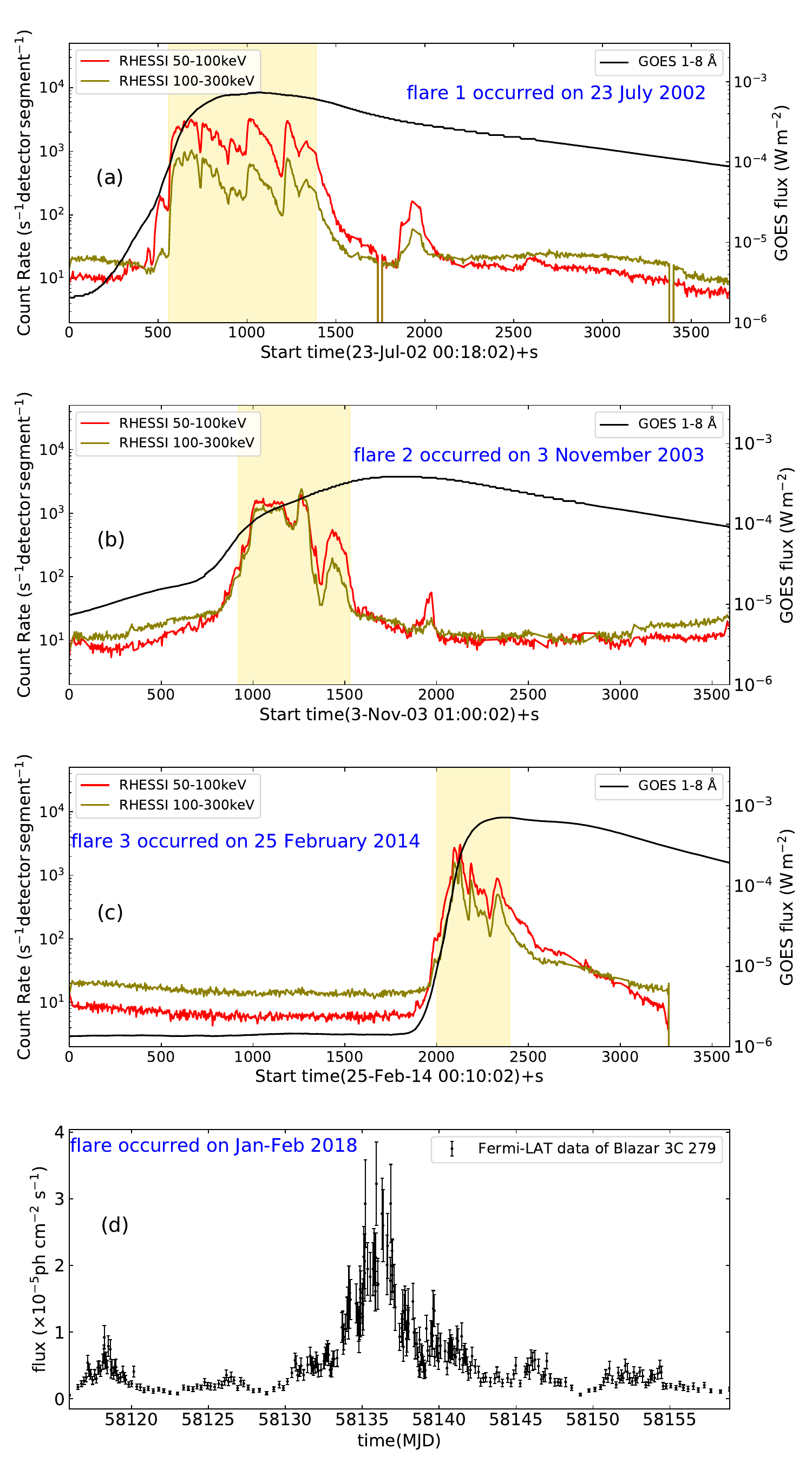}}
    \caption{Comparison of the full-duration GOES soft X-ray and RHESSI hard X-ray profiles of solar flares with the $\gamma$-ray light curves of blazar 3C 279 flaring states for visual context. Panels (a)-(c) show solar flares 1-3, respectively. In each panel, red, yellow, and black curves denote the RHESSI 50-100 keV, 100-300 keV, and GOES 1-8$\text{\AA}$ soft X-ray light curves. The gold shaded region indicates the time interval used in our analysis. The panel (d) is the $\gamma$-ray flare light curve of a typical blazar 3C 279. 
 \label{fig:LC}}
\end{figure}

We select three bright solar hard X-ray flares: the X4.8-class flare on 23 July 2002 (flare 1), the X3.9-class flare on 3 November 2003 (flare 2), and the X4.9-class flare on 25 February 2014 (flare 3), as representative cases for analysis.
They were selected based on the following scientific considerations: (1) Geostationary Operational Environmental Satellites (GOES) class$\ge$X3 (with peak 1-8 $\rm \AA$ soft X-ray flux measured by GOES exceeding $3 \times 10^{-4}\ \mathrm{W/m^2}$), corresponding to strong hard X-ray emission; (2) Continuous, high-quality RHESSI observations covering the entire impulsive phase, ensuring sufficient signal-to-noise for reliable analysis; (3) Pronounced and structured variability in the 50-100 keV and 100-300 keV RHESSI light curves, providing clear temporal signatures for our variability characterization framework. 
Furthermore, they are well-documented benchmark events in RHESSI solar flare studies \citep[e.g.,][]{2003ApJ...595L..97H,2004ApJ...611L..53L,2005A&A...434.1173G,2023CosRe..61..265S}.
To provide a clearer identification of the solar flares analyzed in this study, Figure~\ref{fig:LC} present the full-duration GOES 1-8 $\text{\AA}$ soft X-ray profiles (the black solid line) together with the original RHESSI 50-100 keV (the red solid line) and 100-300 keV (the yellow solid line) hard X-ray profiles for each flare (three solar flares shown in panels (a)(b)(c) respectively), where the time intervals used in our analysis are marked (the gold shaded region).
For visual comparison, a representative $\gamma$-ray flare light curve of blazar 3C 279 is also shown side by side (panel (d)). 
The specific analyzed time intervals for the solar flares are listed in Table~\ref{tab:Fitting results}.

DRW, SHO and Mat$\acute{\rm e}$rn$-3/2$ models are chosen to describe these RHESSI hard X-ray flare light curves.
From the analysis of the fit between the observed data and modeled light curve, the distribution of standard residuals, as well as the autocorrelation functions (ACF) of both the standard residuals and their squares \citep[the goodness-of-fit criteria described in detail in][]{2022ApJ...930..157Z}, no significant differences among the three models for any of the flares.
Further evaluation using the corrected Akaike information criterion ($\rm AIC_{\rm C}$), with values listed in Table~\ref{tab:Fitting results},  shows that the SHO model outperforms the DRW and Mat$\acute{\rm e}$rn$-3/2$ models with $\Delta \rm AIC_{\rm C}>10$ for all flares across both energy bands.

\begin{figure*}
    \centering
    {\includegraphics[width=1\linewidth]{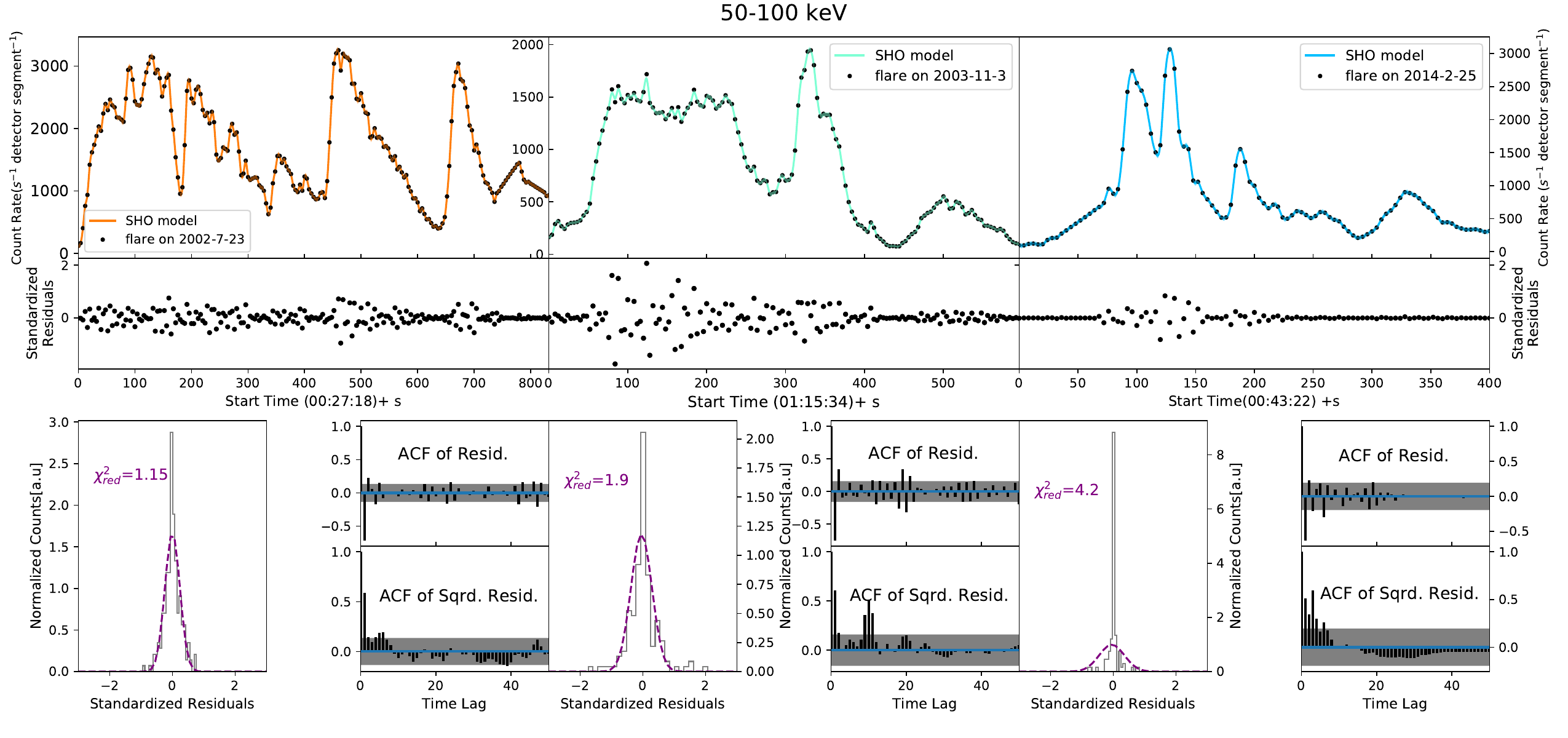}}
    \caption{SHO fitting results for the solar flares in the 50-100 keV energy range. 
    The left, middle and right columns correspond to the fitting results for solar flares occurred on 23 July  2002, 3 November 2003 and 25 February 2014 respectively.
    For each column, we give the RHESSI observed data (black points) and the modeled light curves (orange/green/blue line) in the top panel. 
    In the middle panel, we show the standardized residuals (black points).
    The bottom left panel shows the probability density of standardized residuals (gray histogram) as well as the best-fit normal distribution (purple dotted line). 
    The ACFs of residuals and the ACFs of squared residuals, along with  the 95$\%$ confidence interval for white noise (the gray region) are shown in the two bottom right panels. 
 \label{fig:50-SHO fit}}
\end{figure*}

\begin{figure*}
    \centering
    {\includegraphics[width=1\linewidth]{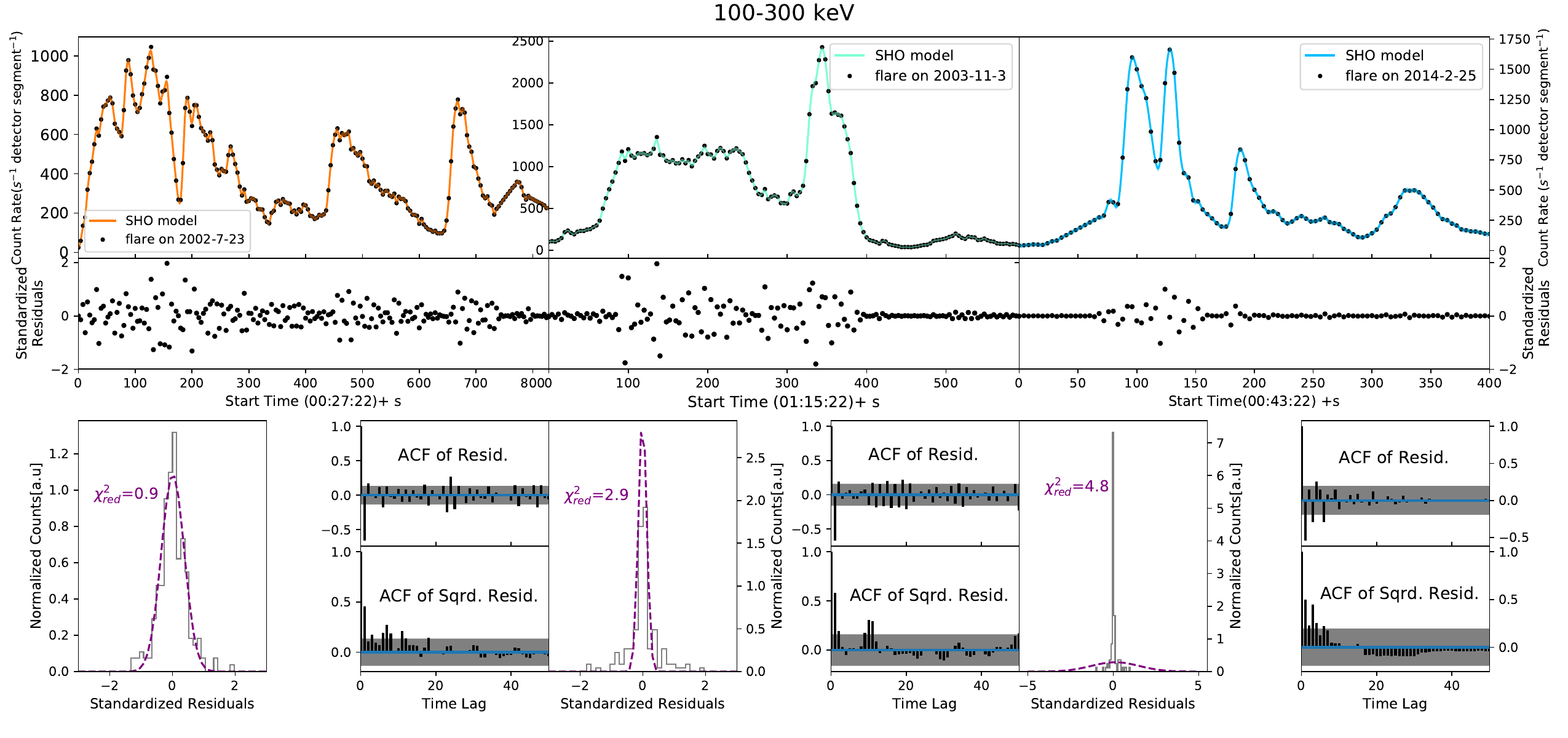}}
    \caption{SHO fitting results for the solar flares in the 100-300 keV energy range. 
    The symbols and lines are the same as those in Figure~\ref{fig:50-SHO fit}.
\label{fig:100-SHO fit}}
\end{figure*}

We present the SHO fitting plots for the solar flares in the 50-100 keV and 100-300 keV energy bands, as shown in Figure~\ref{fig:50-SHO fit} and Figure~\ref{fig:100-SHO fit}.
From left to right, the panels display the results for flare 1, flare 2 and flare 3 respectively.
Although the residual distribution for the flare 3 does not fit a normal distribution well (the bottom left panel for the third columns in Figure~\ref{fig:50-SHO fit} and Figure~\ref{fig:100-SHO fit}), the concentration of residuals around zero and the lack of trends or patterns suggest that the SHO model is still considered valid. 

\begin{figure*}
    \centering
    \begin{minipage}{0.32\textwidth}
      \centering
      \includegraphics[width=1\linewidth]{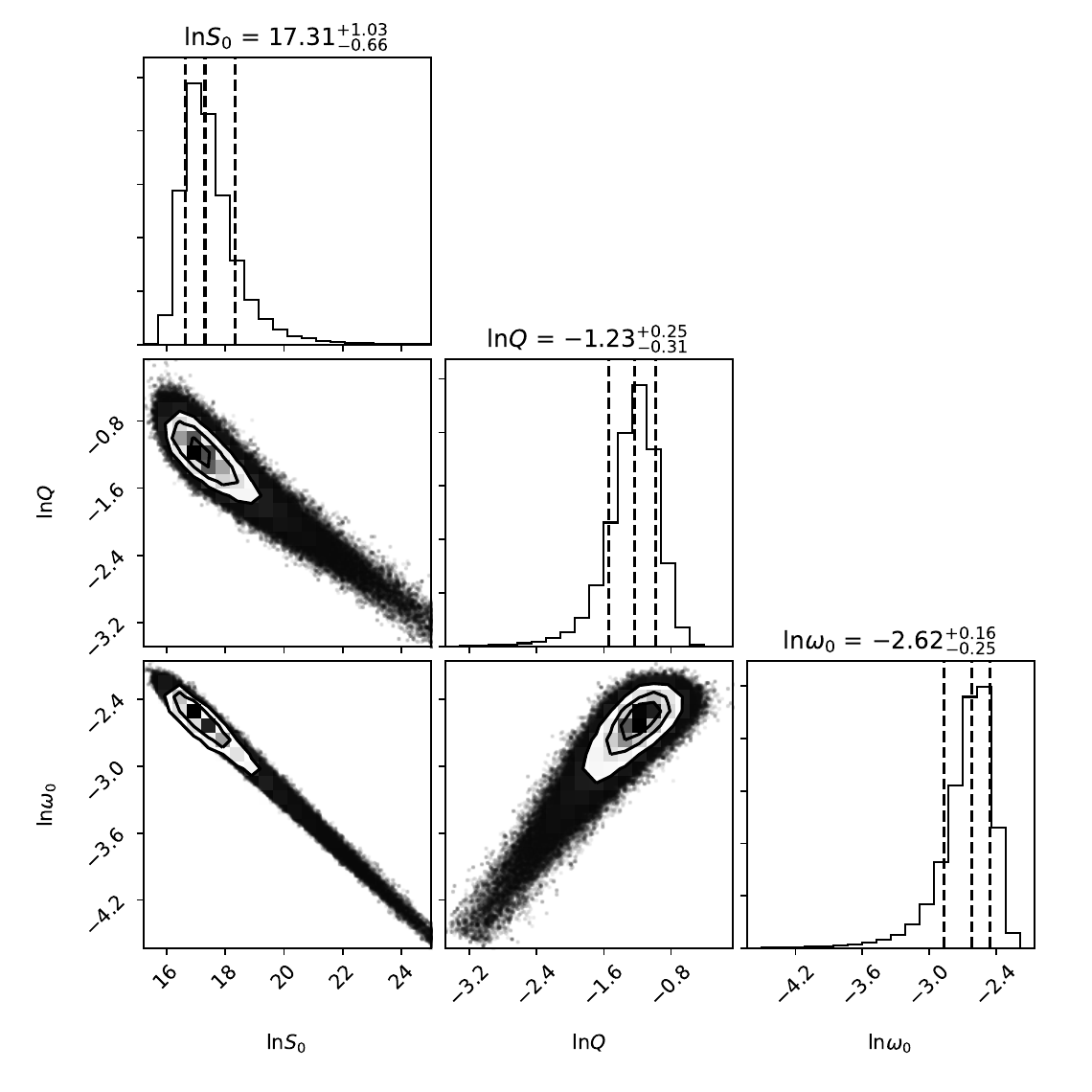}
      
    \end{minipage} \hfill
    \begin{minipage}{0.32\textwidth}
      \centering
      \includegraphics[width=1\linewidth]{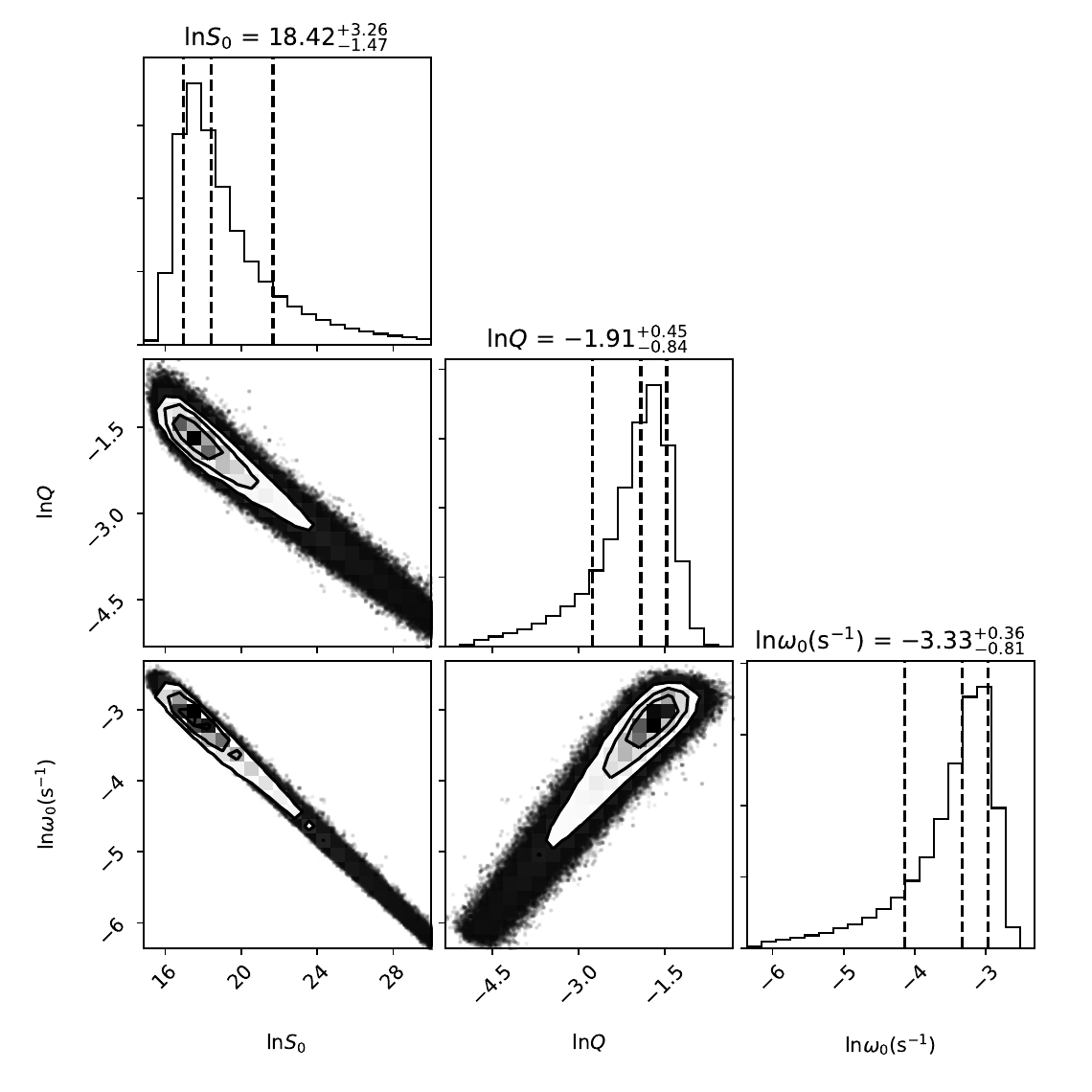}

     \end{minipage} \hfill
    \begin{minipage}{0.33\textwidth}
      \centering
      \includegraphics[width=1\linewidth]{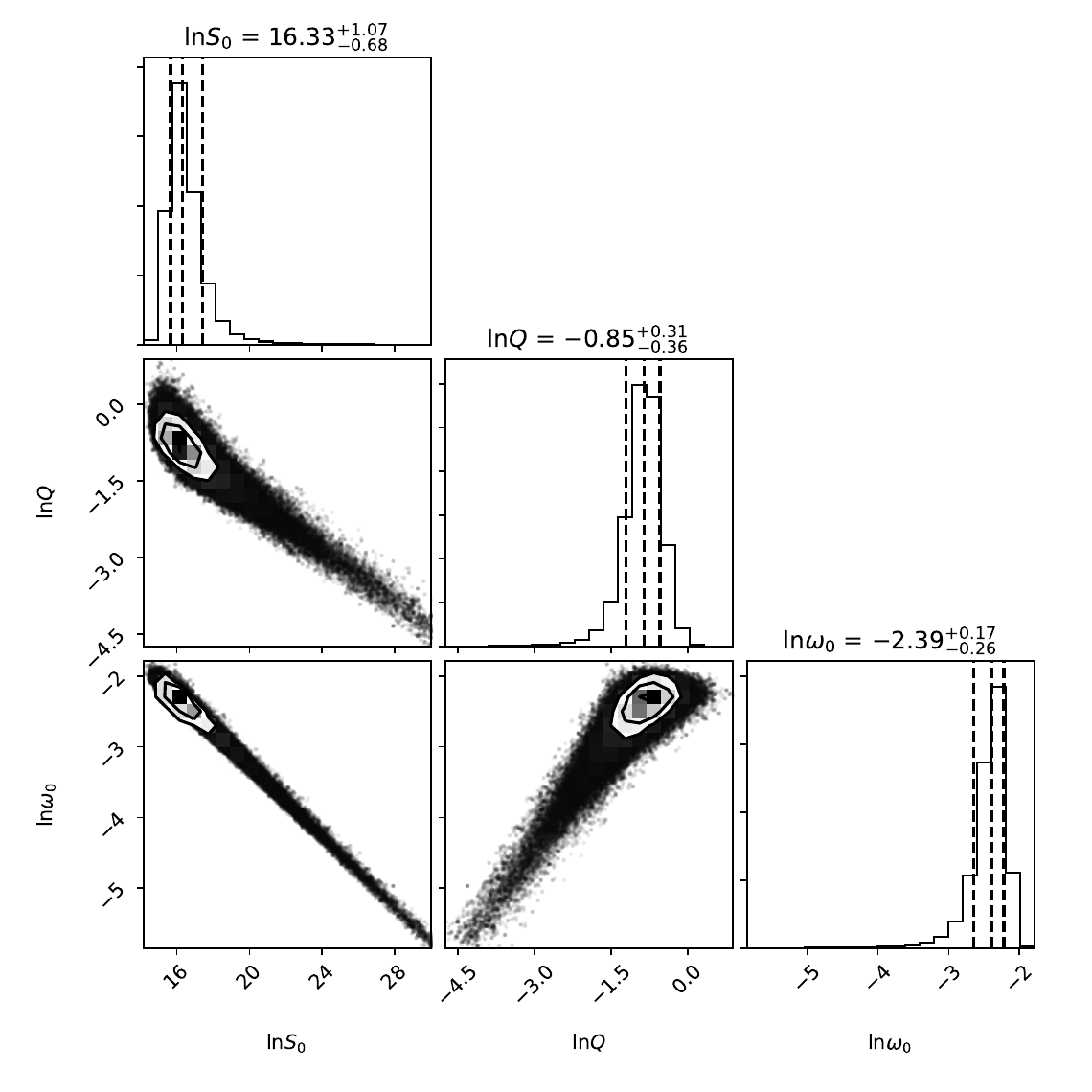}
      
    \end{minipage} \hfill
    \begin{minipage}{0.32\textwidth}
      \centering
      \includegraphics[width=1\linewidth]{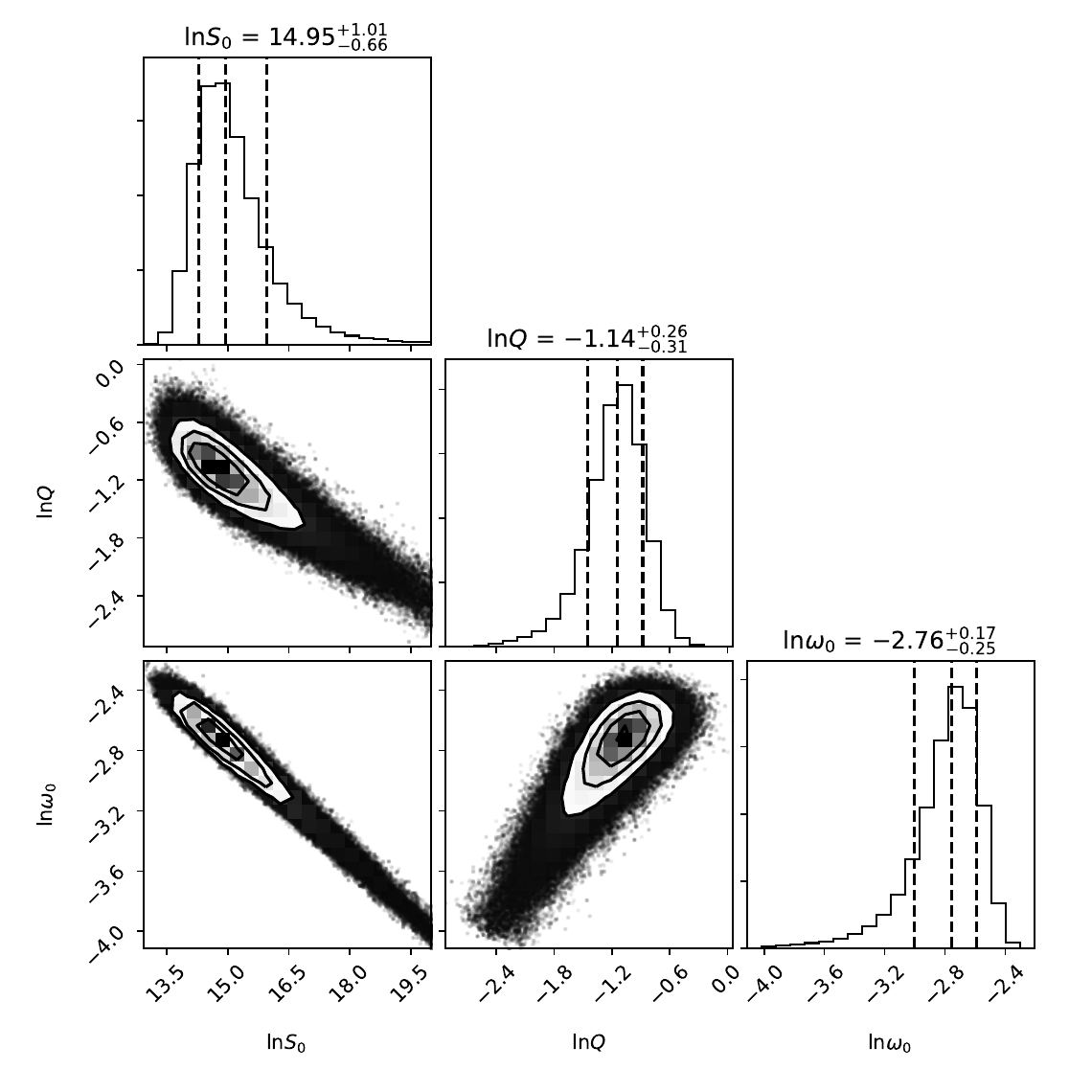}
      
    \end{minipage} \hfill
    \begin{minipage}{0.32\textwidth}
      \centering
      \includegraphics[width=1\linewidth]{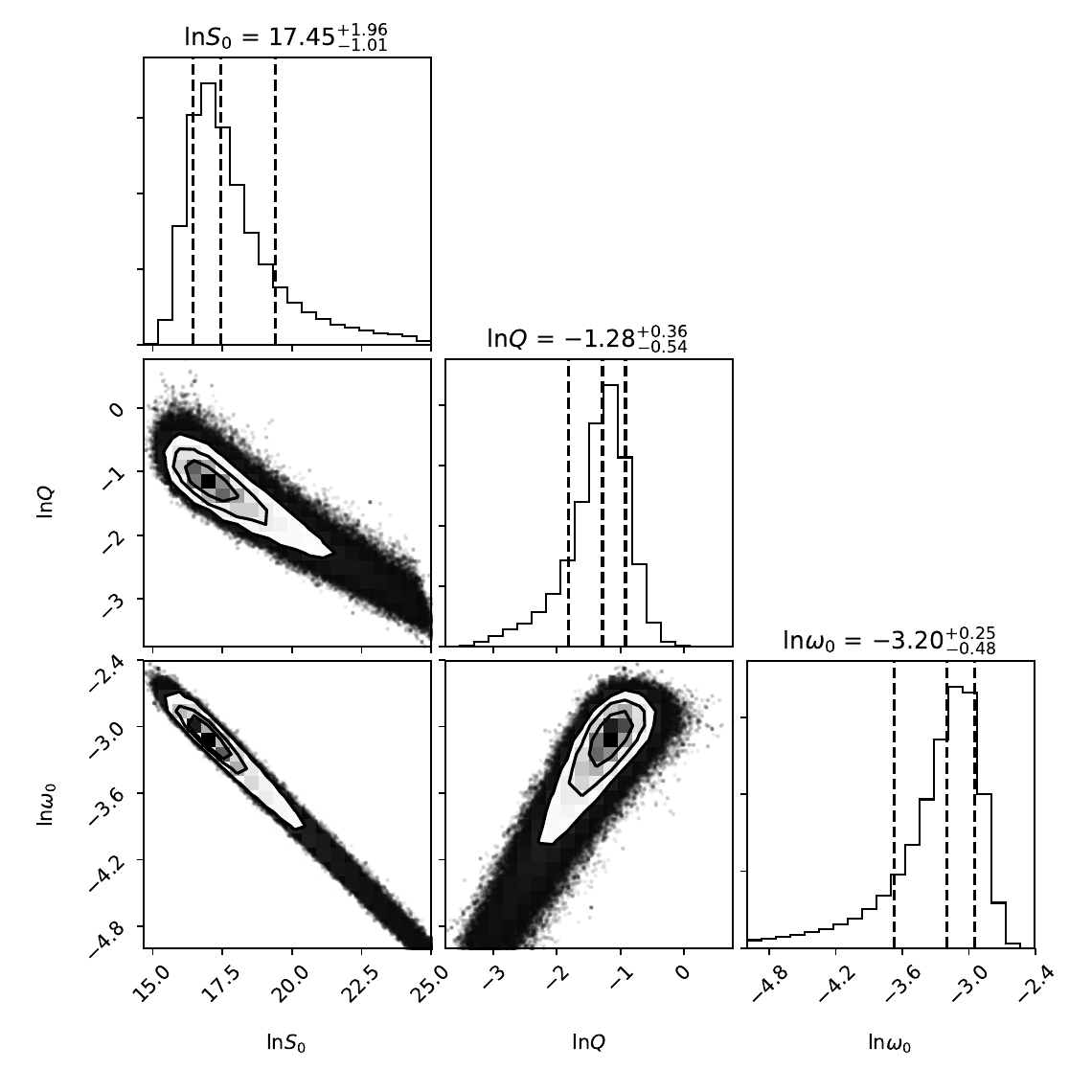}

     \end{minipage} \hfill
    \begin{minipage}{0.32\textwidth}
      \centering
      \includegraphics[width=1\linewidth]{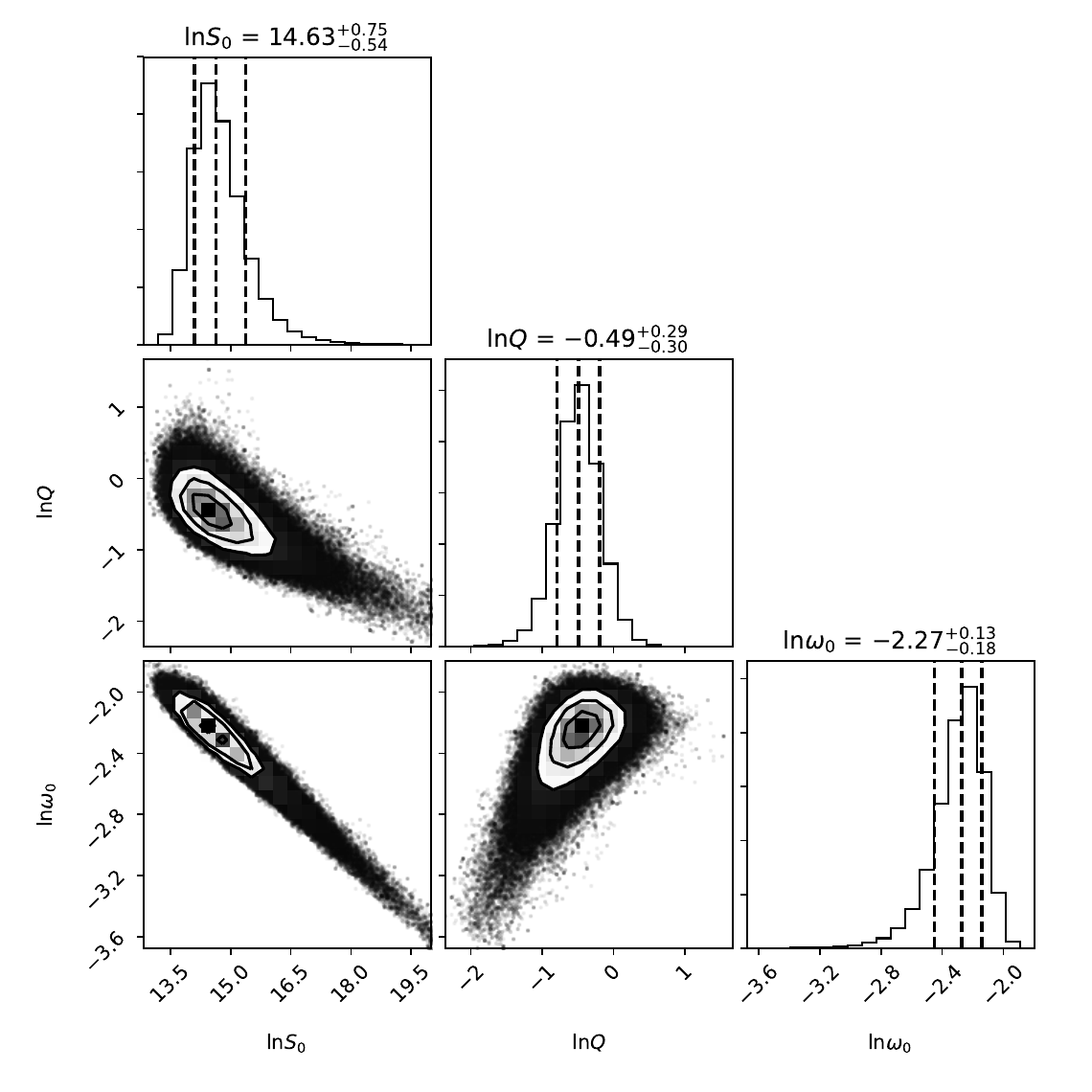}
      
    \end{minipage} 
    \caption{The posterior probability density distribution of the SHO parameters for solar flare 1 (the left column), flare 2 (the middle column) and flare 3 (the right column) in 50-100 keV ( the top row) and 100-300 keV (the bottom row). 
    The vertical dotted lines represent the median value and 68$\%$ confidence intervals of the distribution of the parameters.   
    \label{fig:param}}
\end{figure*}

The corresponding posterior probability density distributions of the SHO parameters are shown in Figure~\ref{fig:param}.
The SHO parameters for all flares are well constrained, with the specific values provided in Table~\ref{tab:Fitting results}.  

\begin{figure*}
    \centering
    {\includegraphics[width=1\linewidth]{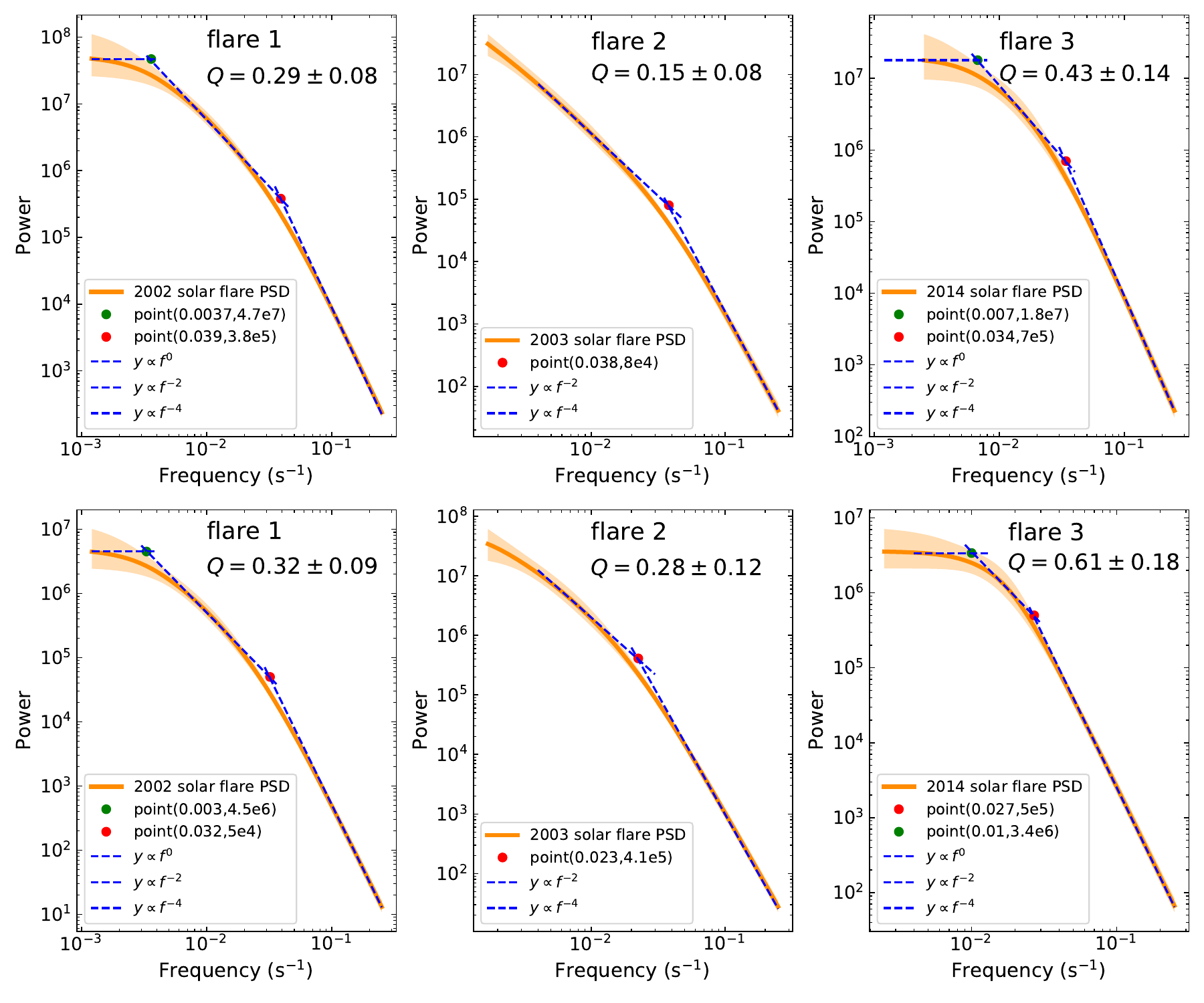}}
    \caption{ PSDs of solar flares in 50-100 keV ( the top row) and 100-300 keV (the bottom row) constructed from the modeling results of the SHO model. The dashed blue lines are reference lines with power-law indexes of $0$, $-2$ and $-4$. 
\label{fig:psd}} 
\end{figure*}

Now we focus on the PSDs shown in Figure~\ref{fig:psd}.
Flare 1 and flare 3 have PSD indexes changing from -4 to -2 and then to 0.
That is, the timescales of $t_{\rm B\underline{~} flat}$ and $t_{\rm B\underline{~} steep}$ are constrained.
The two timescales derived from flare 3 in 100-300 keV are in close proximity to each other.
However, the PSD indexes of flare 2 in two energy bands evolve from -4 to -2, with only $t_{\rm B\underline{~} steep}$ constrained.
The spectral indexes of PSDs and the characteristic timescales of these solar flares are listed in the Table~\ref{tab:information}.
$t_{\rm B\underline{~} steep}$ and $t_{\rm B\underline{~} flat}$ of the solar flares are in the range of 4-7 s and 16-53 s, respectively.
The relationship between these characteristic timescales and the broken frequency $f_{b}$ of the PSD is $t=1/(2\pi f_{b})$.

\begin{deluxetable*}{cccccccccc}
	\tablecaption{Fitting results of solar flares.\label{tab:Fitting results}}
	\tablewidth{30pt}
	\setlength{\tabcolsep}{3mm}{
	\tablehead{
		\colhead{Solar flare} &\colhead{Time period}  & \multicolumn{3}{c}{Parameter of SHO} & \multicolumn{3}{c}{$\rm AIC_{C}$} \\
		\cmidrule(r){3-5} \cmidrule(r){6-8}  
		\colhead{} & \colhead{} & \colhead{ln $S_{0}$} & \colhead{ln Q} & \colhead{ln $\omega_{0}$} & \colhead{SHO} & \colhead{DRW} & \colhead{Mat$\acute{\rm e}$rn$-3/2$}  \\
		\colhead{(1)} & \colhead{(2)} & \colhead{(3)} & \colhead{(4)} & \colhead{(5)} & \colhead{(6)} & \colhead{(7)} & \colhead{(8)} 
	}
	\startdata
	flare 1  &2002-07-23 00:27:18-00:41:10 &$17.31^{+1.03}_{-0.66}$ & $-1.23^{+0.25}_{-0.31}$ & $-2.62^{+0.16}_{-0.26}$ & 2772 & 3461 & 3372 \\
	flare 2  &2003-11-03 01:15:34-01:25:30 & $18.42^{+3.26}_{-1.47}$ & $-1.91^{+0.45}_{-0.84}$ & $-3.33^{+0.36}_{-0.81}$ & 1775 & 1987  & 2321 \\
        flare 3  &2014-02-25 00:43:22-00:50:02 & $16.33^{+1.07}_{-0.68}$ & $-0.85^{+0.31}_{-0.36}$ & $-2.39^{+0.17}_{-0.26}$ & 1346 & 1658  & 1601 \\
  \hline
  flare 1 & 2002-07-23 00:27:22-00:41:10 &$14.95^{+1.01}_{-0.66}$ & $-1.14^{+0.26}_{-0.31}$ & $-2.76^{+0.17}_{-0.25}$ & 2237 & 2352 & 2860 \\
 flare 2  & 2003-11-03 01:15:22-00:25:14 &$17.45^{+1.96}_{-1.01}$ & $-1.28^{+0.36}_{-0.54}$ & $-3.20^{+0.25}_{-0.48}$ & 1758 & 1959 & 2319 \\ 
 flare 3  & 2014-02-25 00:43:22-00:50:02 &$14.63^{+0.75}_{-0.54}$ & $-0.49^{+0.29}_{-0.30}$ & $-2.27^{+0.13}_{-0.18}$ & 1235 & 1414 & 1483
   \enddata
	\tablecomments{ 
	(1) source name, (2) time period of the flare, (3)(4)(5) posterior parameters of modeling light curves with SHO model, (6)(7)(8) $\rm AIC_{C}$ values of SHO, DRW and  Mat$\acute{\rm e}$rn$-3/2$ model. The uncertainties of model parameters represent $1\sigma$ confidence intervals. The upper part shows the results for 50-100 keV, and the lower part for 100-300 keV.
		}}
\end{deluxetable*}

\subsection{Comparison of flares from blazars, magnetars and the Sun}
In previous work, we have analyzed the $\gamma$-ray blazar flares (of 3C 454.3 and 3C 279) and the XRBs occurred in magnetar SGR 1935+2154 by GP.
The light curves of these flares also follow the SHO model.
By using the solar X-ray flare results as a reference, we seek to compare the variability patterns and timescales of blazar $\gamma$-ray flares and XRBs in magnetars, further investigating the physical origins of these distinct flare events.

\begin{figure*}
    \centering
    {\includegraphics[width=1\linewidth]{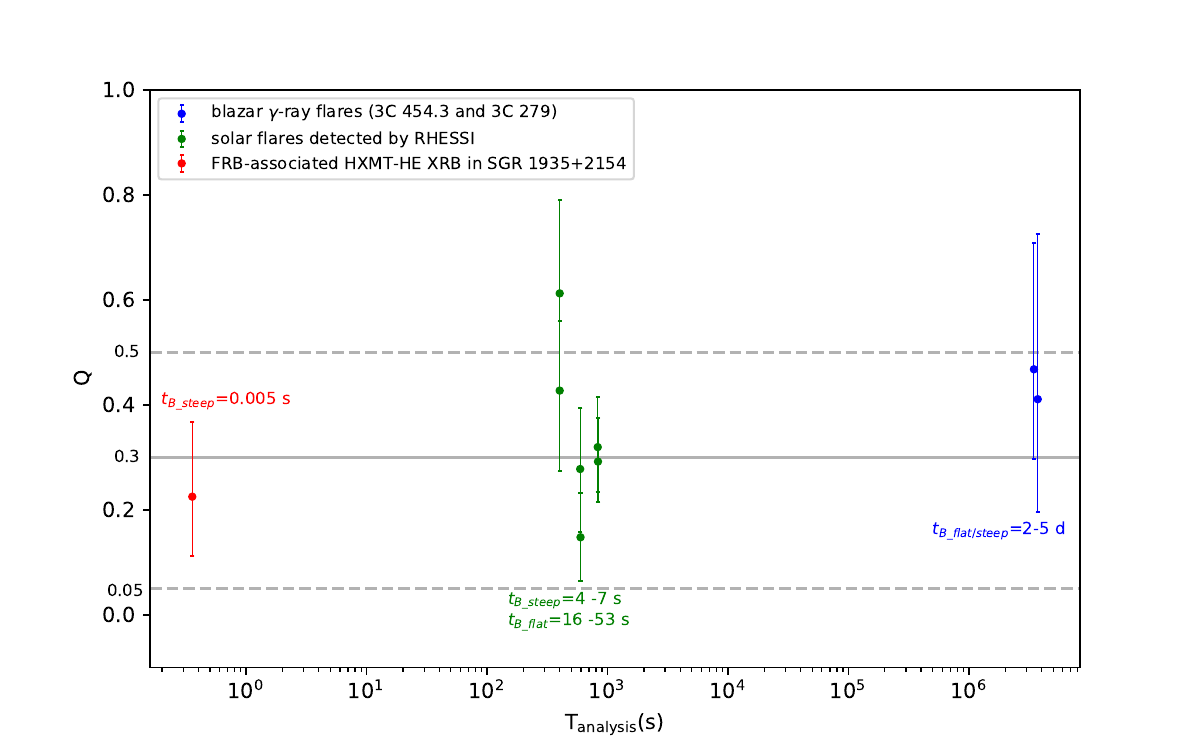}}
    \caption{ Relationship between the parameter $Q$ derived from the SHO model fits and the flare analysis timescale $\rm T_{analysis}$ for different types of sources.}
    \label{fig:Q-T} 
\end{figure*}

Considering that the exact mode of the SHO model is determined by the quality oscillation factor $Q$, we establish a two-dimensional relationship between the parameter $Q$ derived from the SHO model fits and the flare analysis timescale $\rm T_{analysis}$ (i.e., the duration of the flare light curves we analyzed), as shown in  Figure~\ref{fig:Q-T}, to probe the relationship between the dynamics of the different burst events and the their local physical scale of the emission region.
For solar flares (the green part in Figure~\ref{fig:Q-T}),
the $Q$ values of flare 1 and flare 2 are generally near or below 0.3, while flare 3 appears to have $Q$ values significantly larger than 0.3 and develops to larger values (near or larger than 0.5).

There is a unique XRB in SGR 1935+2154 that is associated with FRB 200428. 
Its high-energy (HE) data from Insight-HXMT show the best correspondence with the FRB radio data. 
The $Q$ value of this XRB in HE is approximately 0.3 \citep[as shown by the red mark in Figure~\ref{fig:Q-T};][]{2024ApJ...971...26T}, which is comparable to that of solar X-ray flares 1 $\&$ 2, yet its analysis timescale $\rm T_{analysis}$ is less than that of solar flares by roughly 3 orders of magnitude.
The normal XRBs in SGR 1935+2154 possess $Q$ values spanning a wide range \citep{2024ApJ...971...26T}, without displaying in the figure.

For $\gamma$-ray blazar flares \citep[the blue part in Figure~\ref{fig:Q-T};][]{2025MNRAS.540.3790Z}, there is a difference of 3-4 orders of magnitude for $\rm T_{analysis}$ compared to solar X-ray flares.
They have $Q$ values generally larger than 0.3, which is similar to that of solar flare 3. Their PSDs nearly exhibit a single break structure \citep{{2025MNRAS.540.3790Z}}.

Additionally, we provide a summary of the PSD characteristics and variability timescales for the analyzed flares in Table~\ref{tab:information}.
The FRB-associated HXMT-HE XRB, like solar flares, possesses a PSD structure with an index changing from -4 to -2, but the timescale $t_{\rm B\underline{~} steep}$ of the HXMT-HE XRB is significantly smaller.
As a result of data constraints, both the FRB-associated HXMT-HE XRB and solar flare 2 are missing a PSD structure with an index changing from -2 to 0.
The blazar $\gamma$-ray flares have the longest observed characteristic timescales, with PSD indexes changing from -4 to 0.
 
\begin{figure*}
    \centering
    {\includegraphics[width=1\linewidth]{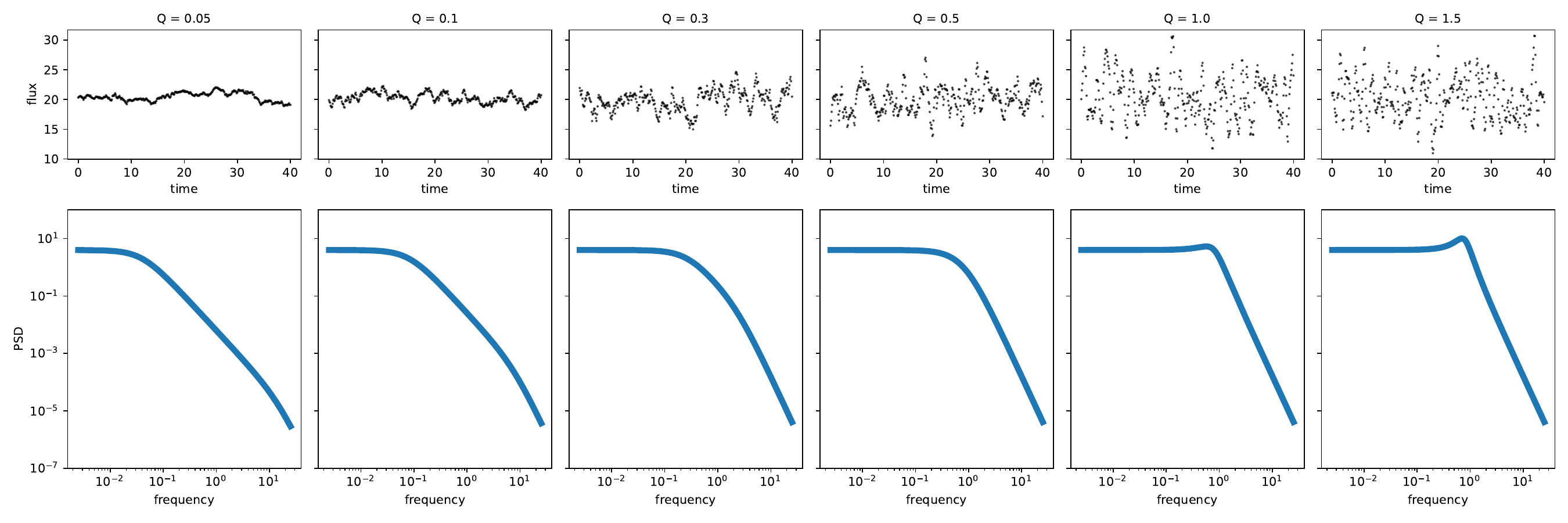}}
    \caption{ Light curves and PSDs simulated by SHO model with different $Q$ values.
    \label{fig:sim_PSD}} 
\end{figure*}

We further simulate the PSDs of the SHO for different $Q$ values, along with the corresponding simulated light curves, as shown in Figure~\ref{fig:sim_PSD}, to better illustrate our results.
The PSD evolution with $Q$ reveals a clear sequence: two distinct breaks at low $Q$ converge into one at $Q=0.5$, after which a QPO peak emerges for $Q>1$. We interpret this entire sequence as reflecting the interaction between two underlying timescales: one from energy injection ($t_{\rm B\underline{~} steep}$; high-frequency break) and the other from energy dissipation ($t_{\rm B\underline{~} flat}$; low-frequency break). The convergence reflects a dynamic rebalancing between the processes, guiding the system to a recovered state of equilibrium, while the eventual QPO marks the excitation of a coherent dynamical mode within the system, likely triggered by the energy injection process.
Through this figure, we can not only better examine how the quality factor $Q$ controls the variability features, such as the amplitude of fluctuations, but also investigate how the two characteristic timescales in the PSD evolve and reflect possible system states. In the context of our solar flare results, Flares 1 and 2 appear to occur in systems where energy is injected relatively rapidly but dissipates more slowly, whereas Flare 3 likely occurs in a near-equilibrium system, where any injected energy is immediately dissipated.

\begin{deluxetable*}{cccc}
\tablecaption{Information of PSDs and timescales of flares.\label{tab:information}}
\tablewidth{0pt}
\tablehead{
\colhead{Source} &\colhead{Type of PSD } &\colhead{Timescale $t_{\rm B\underline{~} steep}$} &\colhead{Timescale $t_{\rm B\underline{~} flat}$}}
 \startdata
the solar X-ray flare 1  & -4$\rightarrow$ -2$\rightarrow$ 0 & 4-5 s & 43-53 s \\
the solar X-ray flare 3 & -4$\rightarrow$ -2$\rightarrow$ 0  & 5-6 s  & 16-23 s  \\
the solar X-ray flare 2 & -4$\rightarrow$ -2 & 4-7 s & $\cdots$ \\
the blazar $\gamma$-ray flares  & -4$\rightarrow$ 0 & 2-5 days & 2-5 days\\ 
HXMT-HE burst associated with FRB 200428 & -4$\rightarrow$ -2 &  0.005 s &  $\cdots$
 \enddata
		
\end{deluxetable*}

\section{Discussion} \label{sec:discussion}
We have analyzed three bright solar X-ray flares that occurred on 23 July 2002, 3 November 2003 and 25 February 2014 using the GP method. 
We fitted the light curves of these flares in the 50-100 keV and 100-300 keV energy bands with the SHO, DRW and Mat$\acute{\rm e}$rn$-3/2$ models. 
The results indicate that the solar X-ray flares prefer the SHO model. 
Both flare 1 and flare 3 constrain two timescales $t_{\rm B\underline{~} flat}$ and $t_{\rm B\underline{~} steep}$ (PSD indexes: -4$\rightarrow$ -2$\rightarrow$ 0), with the two timescales of flare 3 being relatively close to each other.
While for flare 2, due to the limitation of data length, only $t_{\rm B\underline{~} steep}$ is constrained (PSD indexes: -4$\rightarrow$ -2). 
$t_{\rm B\underline{~} steep}$ for the solar X-ray flares are in the range of 4-7 s, and $t_{\rm B\underline{~} flat}$ are in the range of 16-53 s.

The variability of the solar X-ray flares, the blazar $\gamma$-ray flares and the FRB-associated HXMT-HE XRB in SGR 1935+2154 all follow SHO model, yet they exhibit differences in $Q$ values, PSDs, and characteristic timescales as shown in Figure~\ref{fig:Q-T} and Table~\ref{tab:information}.

We aim to investigate the underlying mechanisms of flaring events from different astronomical objects through a comparative framework, leveraging insights from solar X-ray flares fitting results and their associated physical origins.
The driving mechanism of solar X-ray flares is widely attributed to magnetic reconnection, which converts stored magnetic energy into plasma heating and particle acceleration \citep{2008A&ARv..16..155K,2019ASSL..458.....A}.
In the solar X-ray flare system and in the framework of magnetic reconnection, the timescale $t_{\rm B\underline{~} steep}$ (4-7 s) likely corresponds to the energy injection phase, 
and the flat decay timescale $t_{\rm B\underline{~} flat}$ (16-53 s) may reflect subsequent dissipative processes, such as plasma cooling and particle thermalization \citep[e.g.,][]{2012ApJ...754..103B,2015ApJ...802...53A}. 
These timescales should be interpreted with caution.

For solar flares, we note that the high-frequency broken timescale ($\sim$ 4-7s) is broadly consistent with the $\rm Alfv\acute{e}n$ transit time in compact coronal loops, suggesting a connection to rapid magnetohydrodynamics (MHD) energy transport or particle acceleration \citep{2009SSRv..149..119N,2018SSRv..214...45M}.
It also closely matches few-second QPP periods observed in hard X-ray emissions and predicted by MHD simulations \citep{2020STP.....6a...3K,2023BAAS...55c.181I}.
In contrast, the low-frequency broken timescale ($\sim$16-53 s) overlaps with the most commonly reported QPP periods in GOES soft X-rays \citep[$\sim$10-40 s;][]{2020ApJ...895...50H} and some hard X-ray emissions \citep[e.g.,][]{2017MNRAS.471L...6L}. 
This overlap between the timescales we obtained and previously reported QPP periods in solar flares suggests that a common underlying energy release mechanism, such as intermittent/repetitive magnetic reconnection, global MHD oscillatory modes (e.g., kink and sausage modes) in coronal structures, or an associated modulation process may be at work \citep{2005psci.book.....A,2009SSRv..149..119N,2013A&A...558A..76R,2016SoPh..291.3143V,2021SoPh..296..188Z,2023BAAS...55c.181I}.

It is important to emphasize that although the PSD-derived characteristic timescales from our GP analysis overlap in magnitude with QPP oscillation timescales in solar flares, they are not identical. The latter represents explicit oscillatory signals (periodic timescale) that may arise from the underlying energy release dynamics, whereas the former reflect the underlying characteristic timescales of the energy release process itself, even in the absence of a dominant coherent oscillation. In other words, our analysis reveals the fundamental timescales of energy injection and dissipation that govern the system’s behavior, of which QPPs may be one particular manifestation. This provides a complementary perspective on the multi-scale nature of reconnection-driven energy release, but they do not uniquely determine the underlying mechanism. Future high-cadence imaging observations, multi-wavelength spectral diagnostics, and quantitative MHD modeling will be essential for distinguishing between these scenarios.

The discovery of the FRB-associated HXMT-HE XRB in SGR 1935+2154 suggests that a physical process in magnetar activity (such as magnetic reconnection or oscillation modes) simultaneously leads to the generation of both XRBs and FRBs \citep{2020Natur.587...54C,2021NatAs...5..378L}. 
The critically damped state of a system could be the condition triggering this combined event. 
In this case, the critically damped regime of this actual system corresponds to $Q \sim 0.3$, which differs from the theoretical value of $Q = 0.5$.
In the comparative framework for the different types of extreme flares, the FRB-associated HXMT-HE XRB have similar $Q$ values to that of solar flares 1 $\&$ 2, indicating similar dynamic behavior. 
The ultra-short timescales of FRB-associated HXMT-HE XRB in SGR 1935+2154 (0.005 s) are consistent with Alfv$\acute{\rm e}$nic crossing times in magnetar magnetospheres \citep{2015ApJ...815...25L,2020ApJ...900L..21Y}, supporting fast magnetic reconnection triggered by sudden magnetic field reconfiguration. 
This supports a similar energy release process between the two class outbursts.
The difference in characteristic timescale may be attributed to the difference in magnetic field strength and emission scale.

For blazar $\gamma$-ray flares, the central values of $Q$ exhibit an upward shift from 0.3, similar to that of solar X-ray flare 3, indicating that they may share certain common dynamic characteristics within their respective systems.
In addition, a $\sim$13-hour QPO signal detected across $\gamma$-ray flux, optical flux, and linear polarization during the flaring state of BL Lacertae (a blazar) has been reported by \cite{2022Natur.609..265J}, and it strongly support an outburst mechanism, i.e., current-driven kink instabilities \citep[e.g.,][]{2020MNRAS.494.1817D}, that modulates the reconnection process.

The diversity in dynamical behaviors among solar X-ray flares, magnetar XRBs, and blazar flares likely stems from differences in magnetic field geometries (e.g., solar loops vs. relativistic jets) and particle acceleration efficiency.
For instance, the 2-5 day timescales of blazar $\gamma$-ray flares may correspond to the light-crossing time of compact emission regions, where it may be the magnetic reconnection in turbulent jets driving rapid particle acceleration.
While the second-scale solar X-ray flares and millisecond-scale XRBs in magnetars may reflect energy release processes directly. 
These high-energy outburst events may share a fundamental reliance on magnetic energy dissipation, such as magnetic reconnection.

There are several ongoing issues. The first is that in the light curve fitting, although the ACF of the residual squares mostly falls within the white noise range, it does not randomly distribute around zero, particularly for the solar X-ray flare 3. Maybe it needs further analysis.
A second issue is that the SHO model provides the statistically optimal fit under the $\rm AIC_c$ criterion for the flare events analyzed in this study, but its derived physical interpretation must be treated with caution. The inferences about flare physics based on the SHO model’s Q-values and PSD structure should be regarded as exploratory at this stage.
Confirmation of the SHO model’s physical superiority over simpler models and the universality of the inferred Q-values requires analysis of a larger and more diverse set of flare events, which we identify as a key direction for future work.
Furthermore, Our analysis focuses on the near-peak phase (approximate impulsive phase) of the flare. When testing an expanded time interval to cover a larger portion of the light curve, we found that none of the SHO, DRW, Mat$\acute{\rm e}$rn$-3/2$ models, or their combinations can adequately capture the variability across the extended time interval, indicating non-stationary data or the presence of multiple physical regimes.
Alternative approaches such as the changepoints method may be required, which expresses a change from one kernel to another. We are currently developing this approach.

\acknowledgments
D.H.Y acknowledges funding support from the National Natural Science Foundation of China (NSFC) under grant No. 12393852. We thank the support from the Postdoctoral Fellowship Program of China Postdoctoral Science Foundation (CPSF) under Grant Number GZB20230618.

\bibliography{main.bib}{}
\bibliographystyle{aasjournal}

\end{CJK*}
\end{document}